\documentstyle [12pt] {article}

\catcode`@=12
\begin{document}
\thispagestyle{empty}
\begin{flushright} UCRHEP-T153\\October 1995\
\end{flushright}
\vspace{1.0in}
\begin{center}
{\Large \bf Increasing R$_b$ and Decreasing R$_c$\\
with New Heavy Quarks\\}
\vspace{1.5in}
{\bf Ernest Ma\\}
\vspace{0.1in}
{\sl Department of Physics\\}
{\sl University of California\\}
{\sl Riverside, California 92521\\}
\vspace{1.5in}
\end{center}
\begin{abstract}\
If the $b$ and $c$ quarks mix with new heavy quarks of weak isospin
$I_3 = -1$ and 0 respectively, then the
$Z \rightarrow b \bar b ~(c \bar c)$ rate is necessarily greater (smaller)
than that of the standard model.  This may be the reason for the R$_b$
excess and R$_c$ deficit observed at LEP.  A possible consequence of this
scenario is the prospective discovery of a new quark $x$ with the dominant
decay $x \rightarrow ch$, then $h \rightarrow b \bar b$, where $h$ is a
neutral Higgs boson.
\end{abstract}
\newpage
\baselineskip 24pt

It has been known for some time\cite{1} that the experimentally measured
$Z \rightarrow b \bar b ~(c \bar c)$ rate is greater (smaller) than that of
the standard model.  With the recent observation of the top quark\cite{2}
at the Tevatron and more precision data\cite{3} from the four LEP experiments,
the two discrepancies have become even sharper, as summarized below.

\begin{center}
\begin{math}
\begin{array} {|c|c|c|c|} \hline
{}~ & \rm Measurement & \rm SM & \rm Pull \\
\hline
R_b & 0.2219 \pm 0.0017 & 0.2156 & 3.7 \\
R_c & 0.1543 \pm 0.0074 & 0.1724 & -2.5 \\
\hline
\end{array}
\end{math}
\end{center}

\noindent Here $R_b \equiv \Gamma (Z \rightarrow b \bar b)/\Gamma (Z
\rightarrow {\rm hadrons})$, $R_c \equiv \Gamma (Z \rightarrow c \bar c)/
\Gamma (Z \rightarrow {\rm hadrons})$, SM stands for the standard-model fit
with $m_t = 178$ GeV and $m_H = 300$ GeV, and ``pull" is defined as the
difference between measurement and fit in units of the measurement error.
If these results are taken at face value, physics beyond the standard model
is indicated.  Previous attempts in this direction have dealt mostly with
$R_b$.  Its excess has been interpreted as due to one-loop corrections of
the $Z b \bar b$ vextex coming from extensions of the standard model, such
as the two-Higgs-doublet model,\cite{4} or the minimal supersymmetric
standard model,\cite{5} or the $SU(3)^3 \times SU(2)_L \times U(1)_Y$
model.\cite{6}  However, the first two scenarios are in potential conflict
with top quark decay\cite{7} and all three fail to account for the large $R_c$
deficit.

The purpose of this note is to point out that the $R_b$ excess and the $R_c$
deficit are naturally explained by the mixing of the $b$ and $c$ quarks with
new heavy quarks of weak isospin $I_3 = -1$ and 0 respectively.  The idea
is very simple.  Consider first the mixing of the $c$ quark with a new
heavy isosinglet quark $x$ of charge 2/3.\cite{8}  Since both $c_R$ and $x_R$
are singlets, we can define $x_R$ to be that which appears in the
gauge-invariant mass term $\bar x_L x_R$.  We then have both $\bar c_L c_R
\bar \phi^0$ and $\bar c_L x_R \bar \phi^0$ Yukawa terms, where $(\phi^+,
\phi^0)$ is the usual Higgs doublet of the standard model.  As a result,
the mass matrix linking $(\bar c_L, \bar x_L)$ to $(c_R, x_R)$ is given by
\begin{equation}
{\cal M} = \left( \begin{array} {c@{\quad}c} m_c & m_{cx} \\ 0 & M_x
\end{array} \right).
\end{equation}
The $c_L - x_L$ mixing is then $\theta_x \sim m_{cx}/M_x$, whereas the
$c_R - x_R$ mixing is $m_c m_{cx}/M_x^2$ which is certainly negligible.  The
physical $Z \rightarrow c \bar c$ rate becomes proportional to
\begin{eqnarray}
\left[ \left( {1 \over 2} - {2 \over 3} \sin^2 \theta_W \right) \cos^2
\theta_x + \left( -{2 \over 3} \sin^2 \theta_W \right) \sin^2 \theta_x
\right]^2 + \left( -{2 \over 3} \sin^2 \theta_W \right)^2 \nonumber \\
= \left( {1 \over 2} \cos^2 \theta_x - {2 \over 3} \sin^2 \theta_W \right)^2
+ \left( -{2 \over 3} \sin^2 \theta_W \right)^2,
\end{eqnarray}
which is clearly a decreasing function of $\theta_x$ for small $\theta_x$.
Similarly, the physical $Z \rightarrow b \bar b$ rate becomes
proportional to
\begin{eqnarray}
\left[ \left( -{1 \over 2} + {1 \over 3} \sin^2 \theta_W \right) \cos^2
\theta_y + \left( -1 + {1 \over 3} \sin^2 \theta_W \right) \sin^2 \theta_y
\right]^2 + \left( {1 \over 3} \sin^2 \theta_W \right)^2 \nonumber \\
= \left( -{1 \over 2} (1 + \sin^2 \theta_y) + {1 \over 3} \sin^2 \theta_W
\right)^2 + \left( {1 \over 3} \sin^2 \theta_W \right)^2,
\end{eqnarray}
which is clearly an increasing function of $\theta_y$.  To be more precise,
we have assumed an isotriplet $y \equiv (y_1, y_2, y_3)$ of quarks which
transforms as (3; 2/3) under the standard $SU(2) \times U(1)$ with
$Q = I_3 + Y$ in both its left-handed and right-handed projections.  The
extended model is thus anomaly-free and we have a gauge-invariant mass term
$\bar y_{1L} y _{1R} + \bar y_{2L} y_{2R} + \bar y_{3L} y_{3R}$ as well as
the Yukawa term $\bar y_{1R} t'_L \phi^+ + \bar y_{2R} (t'_L \phi^0 +
b_L \phi^+)/\sqrt 2 + \bar y_{3R} b_L \phi^0$, where $t' = V_{tb}^* t +
V_{cb}^* c + V_{ub}^* u$.  Hence $b$ mixes with $y_3$
and $t'$ with $y_2$.  We assume that $M_y > m_t$.

To fit the updated LEP measurements,\cite{3} we need
\begin{eqnarray}
\sin^2 \theta_x &=& 0.045 \pm 0.019, \\
\sin^2 \theta_y &=& 0.0127 \pm 0.0034.
\end{eqnarray}
These numbers are perfectly consistent with the experimentally known entries
of the $3 \times 3$ weak charged-current mixing matrix.\cite{9}  The
precisely measured entries $|V_{ud}|$ and $|V_{us}|$ are not affected.
Others can be reinterpreted without contradiction.  For example, the
experimental value $|V_{cd}|$ may be written as $|V'_{cd}| \cos \theta_x$
and $|V_{cb}|$ as $|V'_{cb}| \cos \theta_x (\cos \theta_y \cos \theta'_y +
\sqrt 2 \sin \theta_y \sin \theta'_y)$, where $\sin \theta'_y \simeq
\sin \theta_y/\sqrt 2$.  In this
notation, $V'$ is again a unitary matrix.

As the result of explaining the experimental values of $R_b$ and $R_c$,
a discrepancy in the total hadronic width is now exposed.  If we keep
$\alpha_s$ at $0.123 \pm 0.006$, then there is a missing $\Delta R$ of
$0.0118 \pm 0.0070$ where the negative correlation between $R_b$ and $R_c$
has been taken into account.  For a smaller value of $\alpha_s$
as indicated in deep-inelastic scattering or the upsilon spectrum or
lattice calculations, the discrepancy would be even worse.  One possible
explanation is that $M_x < M_Z - m_c$ so that $Z$ decays into $c \bar x +
x \bar c$ with a rate proportional to $\sin^2 \theta_x \cos^2 \theta_x/2$.
To obtain $\Delta R > 0.0048$, we would need $M_x < 72$ GeV.  In that case,
$x \bar x$ production at the Tevatron would be plentiful and easily
identifiable unless $x$ decays predominantly into hadrons.  Actually, this
may well happen here because the decay chain $x \rightarrow c h$, then $h
\rightarrow b \bar b$, where $h$ is the standard-model Higgs boson, is
dominant if kinematically allowed, and the existence of the heavy quark $x$
would be hidden at the Tevatron from a search of
its semileptonic decay modes.  Since the present experimental lower bound
of $m_h$ is about 65 GeV (which comes from trying to detect $Z \rightarrow
h + {\rm leptons}$), there is only a narrow window of
opportunity for this scenario to be correct.  On the other hand, if there
are two Higgs doublets, then $h$ is in general a linear combination of
two states, hence the $hZZ$ coupling would be reduced and the experimental
bound on $m_h$ would be lowered accordingly.

If $M_x$ is indeed less than 72 GeV, then it can be confirmed in the near
future at LEP, which will gradually step up in energy to about 190 GeV.
The $e^- e^+ \rightarrow x \bar x$ cross section (not including radiative
corrections) is given by
\begin{equation}
\sigma = {{8 \pi \alpha^2} \over {9 s}} \sqrt {1 - {{4 M_x^2} \over s}}
\left( 1 + {{2 M_x^2} \over s} \right) \left\{ \left| 1 - {{s(1 - 2 \sin^2
\theta_W)} \over {2 \cos^2 \theta_W (s - M_Z^2 + i M_Z \Gamma_Z)}} \right|^2
+ \left| 1 + {{s \tan^2 \theta_W} \over {s - M_Z^2 + i M_Z \Gamma_Z}}
\right|^2 \right\},
\end{equation}
which is about 4~pb at $\sqrt s = 160$ GeV for $M_x = 70$ GeV.  This
increase in the hadronic rate should be detectable across the $x \bar x$
threshold.  The decay of $x$ will be dominantly into $ch$, then $h
\rightarrow b \bar b$, as discussed in the previous paragraph.  Such a
signature should be easily identifiable at LEP2.

With $c-x$ and $b-y$ mixing, the forward-backward asymmetries of $c \bar c$
and $b \bar b$ production at LEP are also affected.  Taking the central value
$\sin^2 \theta_x = 0.045$, the predicted value of $A^c_{FB}$ is about 6\%
below that of the standard model.

\begin{center}
\begin{math}
\begin{array} {|c|c|c|c|} \hline
g_V^c & g_A^c & A_{FB}^c & A_{FB}^c ({\rm LEP}) \\
\hline
0.1685 & 0.4775 & 0.0685 & 0.0725 \pm 0.0058 \\
\hline
\end{array}
\end{math}
\end{center}

In the case of $A_{FB}^b$, taking the central value $\sin^2 \theta_y =
0.0127$, its predicted value is only about 0.2\% above that of the
standard model.

\begin{center}
\begin{math}
\begin{array} {|c|c|c|c|} \hline
g_V^b & g_A^b & A_{FB}^b & A_{FB}^b ({\rm LEP}) \\
\hline
-0.3519 & -0.5064 & 0.1022 & 0.0999 \pm 0.0017 \\
\hline
\end{array}
\end{math}
\end{center}

\noindent It is seen that both asymmetries agree well with the experimental
measurements.

Tree-level flavor-changing neutral-current (FCNC) effects are present in this
model. It has been assumed that the new quarks $x$, $y_3$, and $y_2$ mix
only with $c$, $b$, and $t'$ respectively.  Hence there is necessarily a
contribution to $D^0 - \bar D^0$ mixing from the interaction
\begin{equation}
{\cal H}_{int} = {{-g} \over {2 \cos \theta_W}} \cos \theta_x \sin^2 \theta'_y
Z_\mu (V'_{ub} {V'_{cb}}^* \bar u_L \gamma^\mu c_L + {V'_{ub}}^* V'_{cb}
\bar c_L \gamma^\mu u_L),
\end{equation}
which results in a value of $\Delta m_D/m_D \sim 10^{-18}$, well
below the experimental bound of $7 \times 10^{-14}$.\cite{9}  In the above,
we have used the central values given in Eqs.~(4) and (5) as well as
$|V_{cb}| = 0.040$, $|V_{ub}/V_{cb}| = 0.08$, and $f_D = 200$ MeV.  Note that
if $d$ and $s$ also mix with $y_3$, then there would be also tree-level FCNC
contributions to $K - \bar K$ and $B - \bar B$ mixing.

There will be a definite
impact on planned $B$ physics measurements.  The famous unitarity triangle
based on the standard-model condition
\begin{equation}
V_{ud}^* V_{ub} + V_{cd}^* V_{cb} + V_{td}^* V_{tb} = 0
\end{equation}
will be modified to read
\begin{equation}
V_{ud}^* V_{ub} + V_{cd}^* V_{cb} / \cos^2 \theta_x + V_{td}^* V_{tb} = 0.
\end{equation}
The oblique radiative corrections $S$, $T$, and $U$ are affected only to the
extent that the new heavy quarks $x$ and $y$ mix with the usual ones.  Since
the mixings are small, these changes are much smaller than the experimental
uncertainties.

In conclusion, it has been suggested in this note that if both the $R_b$
excess and the $R_c$ deficit at LEP are due to new physics, a simple
explanation is that the $b$ and $c$ quarks mix with new heavy quarks of
weak isospin $I_3 = -1$ and 0 respectively.  To keep the total hadronic
rate from $Z$ decay at about the standard-model level which does agree
with data, the new quark $x$ may have to be light enough so that $Z \rightarrow
c \bar x + x \bar c$ is possible at LEP, and $e^- e^+ \rightarrow x \bar x$
possible at LEP2.  For $x$ to have evaded detection at the Tevatron, it must
decay dominantly into hadrons.  In this scenario, that means $x \rightarrow
ch$, where $h$ is a neutral Higgs boson which then decays into $b \bar b$.
This may be detectable already at LEP from $Z \rightarrow c \bar x + x \bar c$
because its branching fraction has to be greater than about $3 \times 10^{-3}$
and should rise above the expected QCD background.  Of course, there may be
other decay modes such as $x \rightarrow s h^+$, where $h^+$ is a charged
Higgs boson which then decays into $c \bar s$ or $\nu_\tau \tau^+$.  The
signal would then be diluted. In any case, the production and detection
of $x \bar x$ at LEP2 would not be a problem if
kinematically allowed.
\vspace{0.3in}
\begin{center} {ACKNOWLEDGEMENT}
\end{center}

I thank Roger Phillips for discussions and an important suggestion.  I also
thank Vernon Barger for correspondence and for reading the manuscript.  This
work was supported in part by the U. S. Department of Energy under Grant
No. DE-FG03-94ER40837.

\newpage
\bibliographystyle{unsrt}

\end{document}